# Electric field gradient wave (EFGW) in iron-based superconductor Ba$_{0.6}$K$_{0.4}$Fe$_2$As$_2$ studied by Mössbauer spectroscopy

A. K. Jasek[1], K. Komędera[1], A. Błachowski[1], K. Ruebenbauer[1*], Z. Bukowski[2], J. G. Storey[3,4], and J. Karpinski[5,6]

[1]Mössbauer Spectroscopy Division, Institute of Physics, Pedagogical University
*ul. Podchorążych 2, PL-30-084 Kraków, Poland*

[2]Institute of Low Temperature and Structure Research, Polish Academy of Sciences
*ul. Okólna 2, PL-50-422 Wrocław, Poland*

[3]Cavendish Laboratory, University of Cambridge
*CB3 0HE, United Kingdom*

[4]School of Chemical and Physical Sciences, Victoria University
*P.O. Box 600, Wellington, New Zealand*

[5]Laboratory for Solid State Physics, ETH Zurich
*CH-8093 Zurich, Switzerland*

[6]Institute of Condensed Matter Physics, EPFL
*CH-1015 Lausanne, Switzerland*

[*]Corresponding author: sfrueben@cyf-kr.edu.pl

**Abstract**

The optimally doped '122' iron-based superconductor Ba$_{0.6}$K$_{0.4}$Fe$_2$As$_2$ has been studied by $^{57}$Fe Mössbauer spectroscopy versus temperature ranging from 4.2 K till 300 K with particular attention paid to the superconducting transition around 38 K. The spectra do not contain magnetic components and they exhibit *quasi*-continuous distribution of quadrupole split doublets. A distribution follows the electric field gradient (EFG) spatial modulation (wave) – EFGW. The EFGW is accompanied by some charge density wave (CDW) having about an order of magnitude lesser influence on the spectrum. The EFGW could be modeled as widely separated narrow sheets with the EFG increasing from small till maximum value almost linearly and subsequently dropping back to the original value in a similar fashion – across the sheet. One encounters very small and almost constant EFG between sheets. The EFGW shape and amplitude as well as the amplitude of CDW are strongly affected by a superconducting transition. All modulations are damped significantly at transition (38 K) and recover at a temperature being about 14 K lower. The maximum quadrupole splitting at 4.2 K amounts to about 2.1 mm/s, while the dispersion of CDW seen on the iron nuclei could be estimated far away from the superconducting gap opening and at low temperature as 0.5 el./a.u.$^3$. It drops to about 0.3 el./a.u.$^3$ just below transition to the superconducting state.

## 1. Introduction

There are several reports on the sensitivity of Mössbauer spectroscopy to the superconducting transition [1-7] in particular for the iron-based superconductors [8, 9]. Variation in the lattice dynamics due to the superconducting transition has been predicted as well [10]. Generally some variation in the recoilless fraction is observed across the transition and is sometimes accompanied by variation in the second order Doppler shift (SOD), but many reports are inconsistent. Some changes of the lattice stiffness have been observed e.g. by the neutron scattering in iron pnictides [11-13]. Mössbauer spectroscopy is generally insensitive to the superconducting transition in classical superconductors [14, 15]. However, for unconventional superconductors where some very short-range pairing mechanisms could apply everything depends on the coherence length of the composite boson (Cooper pair) as long as local effects are considered. One has to note that superconductivity drastically modifies density of the electronic states at the Fermi surface, and the latter has influence on the hyperfine interactions. Hence, a search by the Mössbauer spectroscopy is justified – a method does not perturbing superconducting state.

We have chosen an optimally doped iron-based superconductor of the '122' family – namely $Ba_{0.6}K_{0.4}Fe_2As_2$. It has a relatively high critical temperature of $T_{sc} = 38$ K [16] and the magnetism is completely suppressed at optimal doping making analysis of the spectra much easier [17]. High-quality samples and large single crystals are available for the '122' family in contrast to other families of iron-based superconductors. One can observe for the sample in question hyperfine parameters, i.e. the quadrupole splitting and the total shift. Some auxiliary information is contained in the absorber line width. The area under the absorption cross-section monitors the recoilless fraction on the resonant atoms. Hence, one can look at the variation of the above parameters across a transition to the superconducting state.

## 2. Experimental

The polycrystalline sample of $Ba_{0.6}K_{0.4}Fe_2As_2$ was prepared by a solid state reaction method from high-purity Ba, K, As, and Fe with natural isotopic composition, as described in Ref. [18].

The Mössbauer absorber was prepared in powder form by mixing 39 mg of $Ba_{0.6}K_{0.4}Fe_2As_2$ with the $B_4C$ carrier. The thickness of the absorber amounted to 19 mg/cm$^2$ of $Ba_{0.6}K_{0.4}Fe_2As_2$. A commercial $^{57}$Co(Rh) source kept at room temperature was applied. A Janis Research Co. SVT-400 cryostat was used to maintain the absorber temperature, with the long time accuracy better than 0.01 K (except at 4.2 K, where the accuracy is better than 0.1 K). A RENON MsAa-3 Mössbauer spectrometer equipped with a Kr-filled proportional counter was used to collect spectra in the photo-peak window. The geometry, count-rate and single channel analyzer window borders were kept constant during all measurements constituting a single uninterrupted series with increasing subsequent temperatures within the range 4.2 – 65 K. Additional spectra were collected at 80 K and 300 K. The velocity scale was calibrated by a He-Ne laser-equipped interferometer. The data were processed within the transmission integral approximation by the Mosgraf-2009 software suite applying GmfpQDW application [19].

## 3. Theoretical background for EFGW and data evaluation method

The Mössbauer spectroscopy is sensitive to the charge (electron) distribution around the resonant nucleus via the isomer shift and the electric quadrupole interaction. The former is possible to observe due to the fact that two nuclear levels are involved and it amounts to $S_I = \alpha(\rho - \rho_S)$, where the parameter $\alpha$ is the so-called calibration constant, while the symbol $\rho$ stands for the electron density on the resonant nucleus in the absorber. The symbol $\rho_S$ denotes corresponding electron density in the source or reference material (constant). For a resonant transition from the ground to the first excited nuclear state of $^{57}$Fe one has $\alpha = -0.291 \, (\text{mm/s}) (\text{a.u.})^3 \, \text{el.}^{-1}$ [20]. A total spectral shift versus some reference material like α-Fe (at normal conditions) or source involves a second order Doppler shift (SOD) $S_D$ as well. However, the latter shift is usually the same for all resonant atoms at a given temperature and pressure provided the source is kept under constant temperature and pressure as well. Hence, a total shift versus reference material (α-Fe at room temperature and normal pressure in the present work) amounts to $S = S_D + S_I$. The electric quadrupole interaction affects solely the first excited nuclear state for aforementioned transition as the ground state has nuclear spin $I_g = 1/2$. The first excited state has spin $I_e = 3/2$ and hence, a doublet is observed with the splitting $\Delta = 2|\varepsilon|$. For isotropic recoilless fraction and completely random absorber this is symmetrical doublet composed of two Lorentzian lines having the same line width $\Gamma$. Note that aforementioned resonant transition is of the pure M1 character. It is assumed that the source is resonantly thin and emits unpolarized radiation as a single Lorentzian line having width $\Gamma_S$. The parameter $\varepsilon$ (quadrupole coupling constant) evaluates to $\varepsilon = [(ecQ_e)/(4E_0)]V_{zz}(1 + \eta^2/3)^{1/2}$. The symbol $e$ stands for the positive elementary charge, while the symbol $c$ denotes speed of light in vacuum. The spectroscopic electric quadrupole moment amounts to $Q_e = +0.17 \, \text{b}$ for the first excited state in $^{57}$Fe [20]. The symbol $E_0$ denotes energy of the resonant transition (14.41-keV), while the symbol $V_{zz}$ stands for the principal component of the electric field gradient tensor (EFG) on the resonant nucleus. The parameter $0 \leq \eta \leq 1$ is the so-called asymmetry parameter of the EFG. It equals null for the axially symmetric EFG. One can measure only the splitting $\Delta$ for a transition above mentioned in the absence of magnetic hyperfine interactions, and for the material being isotropic in the sense defined above.

The charge density wave (CDW) is a spatial modulation of the charge (electron) density and for three dimensional or layered systems it is usually approximated by the time independent standing plane wave with the spatial period quite often being incommensurate with the lattice period in the same direction. The s electrons in CDW affect the isomer shift on resonant nuclei leading to the distribution of the isomer shifts. A contribution from the minor relativistic p electrons could be neglected for such light atoms like iron. Cieślak and Dubiel [21] performed detailed studies on the influence of the CDW shape on the Mössbauer spectra. For a similar modulation of the density of electrons with higher angular momentum than zero one can expect modulation of the EFG in addition to the constant EFG induced locally by some symmetry breaking below cubic. The latter effect could be much stronger than the previous one, i.e. the isomer shift modulation, (about an order of magnitude) due to the local enhancement caused by redistribution of the valence electrons. Hence, the parameter $\varepsilon$ could be written in the form $\varepsilon = \varepsilon(\mathbf{q} \bullet \mathbf{r}) = [(ecQ_e)/(4E_0)]V_{zz}(\mathbf{q} \bullet \mathbf{r})[1 + \eta^2(\mathbf{q} \bullet \mathbf{r})/3]^{1/2}$. The symbol $\mathbf{q}$ denotes wave vector of the time independent standing wave leading to the modulation, while the symbol **r** stands for a position of the particular resonant nucleus. The latter type of modulation is abbreviated further as EFGW (electric field gradient wave). One cannot fit simultaneously CDW and EFGW shapes due to the limited resolution. The parameter $\varepsilon = \varepsilon(\mathbf{q} \bullet \mathbf{r})$ could be expanded into harmonics as follows:

$$\varepsilon = \varepsilon(\mathbf{q} \bullet \mathbf{r}) = \varepsilon_0 + \sum_{n=1}^{N} \left( a_n \cos[n(\mathbf{q} \bullet \mathbf{r})] + b_n \sin[n(\mathbf{q} \bullet \mathbf{r})] \right).$$

(1)

The symbol $\varepsilon_0$ stands for a constant component. The parameters $a_n$ and $b_n$ denote amplitudes of subsequent harmonics. For a complex shape of EFGW ($N \gg 1$) it is virtually impossible to fit independent amplitudes of various harmonics due to the limited resolution. The situation is much better in the case of the spin density waves (SDW) as the resolution of the magnetically split spectra is much higher [22]. Hence, some approximation is necessary in the case of EFGW (and even more in the case of CDW). We have used the following approximation within the range $0 \leq \mathbf{q} \bullet \mathbf{r} \leq 2\pi$:

$$\varepsilon = \varepsilon(\mathbf{q} \bullet \mathbf{r}) = \varepsilon_0 + A F_{max}^{-1} F(\mathbf{q} \bullet \mathbf{r}).$$

(2)

Here the symbol $A > 0$ stands for the amplitude of the modulation, while the symbol $F_{max} > 0$ denotes maximum value of the function $F(\mathbf{q} \bullet \mathbf{r})$ taking on the following form:

$$F(\mathbf{q} \bullet \mathbf{r}) = \sin(\mathbf{q} \bullet \mathbf{r}) \left\{ \exp\left[ -\beta^2 \left( \frac{\mathbf{q} \bullet \mathbf{r}}{2\pi} - \frac{1}{4} \right)^2 \right] + \exp\left[ -\beta^2 \left( \frac{\mathbf{q} \bullet \mathbf{r}}{2\pi} - \frac{3}{4} \right)^2 \right] \right\}.$$

(3)

The shape of EFGW is described by the adjustable parameter $\beta$. This approximation works reasonably and it relies on the two adjustable parameters only $A$ and $\beta$ being therefore numerically stable. Hence, the absorption cross-section is described by a *quasi*-continuous set of symmetrical doublets having common total shift $S$ (average total shift) and being composed of Lorentzians having all the same line width. The absorber dimensionless resonant thickness $t_A$ is an adjustable parameter within standard transmission integral used to fit the spectrum. Another parameter describing transmission integral is the effective source recoilless fraction, i.e. a recoilless fraction of the source corrected for the detector background under the resonant γ-ray line. However, for a single series of uninterrupted measurements with approximately constant average count-rate (within the linear amplitude and frequency response range of the detector system) above parameter could be kept constant upon having measured it independently. It has been set here to $f_S / \lambda = 0.56$ with the symbol $f_S$ denoting recoilless fraction of the source and symbol $\lambda > 1$ standing for the background counts correction. The parameter $\lambda$ is defined as $\lambda = (s+b)/s$, where $s$ stands for the number of counts due to the resonant line (both recoilless and with recoil), while $b$ denotes number of counts belonging to the background. Both numbers of counts are those accepted within the window of the analyzer.

For large values of the parameter $\beta$ ($\beta \gg 0$) one obtains the following approximate distribution of the splitting parameter $\rho(\Delta) = C\delta(\Delta - \Delta_0) + [(1-C)/(\Delta_{max} - \Delta_0)]$ with $\Delta_0 \leq \Delta \leq \Delta_{max}$, where $0 \leq \Delta_0 \ll \Delta_{max}$. The parameter $0 \leq C \leq 1$ accounts for the contribution of the "narrow" component as the symbol $\delta(\Delta - \Delta_0)$ denotes Dirac delta function. The maximum splitting satisfies the following condition $\Delta_{max} = 2(|\varepsilon_0| + A)$. Corresponding distribution expressed in terms of the parameter $\varepsilon - \varepsilon_0$ takes on the form $\rho(\varepsilon - \varepsilon_0) = C\delta(\varepsilon - \varepsilon_0) + [(1-C)/(2\varepsilon_{max})]$ within the range $-\varepsilon_{max} \leq \varepsilon - \varepsilon_0 \leq \varepsilon_{max}$ and for $\varepsilon_{max} = \frac{1}{2}(\Delta_{max} - \Delta_0) = A > 0$. Hence, one can conclude that the spectrum is sensitive under above conditions to the EFGW in the one quarter of the period (first quarter, i.e. for $0 \leq \mathbf{q} \bullet \mathbf{r} \leq \pi/2$) and information about the sign of the principal EFG component is entirely lost. Such shape of the distribution is an indication that the EFGW varies almost linearly within some narrow range of the phase space – going up to the extremum value and falling back to the background. The EFGW remains small and almost constant between narrow regions of strong variability above mentioned. One has to note that eventual "rotation" (described in general by three Eulerian angles) of the total EFG along the propagation direction remains undetectable in the present context.

A signature of the CDW accompanying much strongly exposed EFGW could be seen in the lowest order as the excess of the absorber line width. Hence, one can estimate variation (dispersion) of the electron density on the resonant nuclei (around the average value) caused by existing CDW according to the following expression $\Delta\rho = \sqrt{(\Gamma^2 - \Gamma_{exp}^2)/\alpha^2}$ due to the incoherent character of the broadening. Here, the symbol $0 < \Gamma_{exp} < \Gamma$ denotes unbroadened line width being slightly larger than the natural line width $\Gamma_0$. In principle, the natural line width $\Gamma_0$ is affected by CDW via the variation of the total (internal) conversion coefficient. However, the latter effect is extremely small and could be safely neglected even for large conversion coefficients like for the resonant transition considered here. We have used the following values $\Gamma_{exp} = \Gamma_S = 0.1$ mm/s, while the natural line width amounts to $\Gamma_0 = 0.097$ mm/s for a transition in question. The last approximation does not account, of course, for the shape of CDW. The shape of CDW cannot be reliably extracted due to the limited resolution and much stronger effect of EFGW. In principle, equation (1) could be used to describe shape of CDW provided the constant $\varepsilon_0$ is replaced by the constant $S$. Usually the parameters $a_n$ and $b_n$ take on different values for CDW and EFGW, respectively. Dispersion $\Delta\rho$ could be expressed as $\Delta\rho = \left\{(2\alpha^2)^{-1} \sum_{n=1}^{N} (a_n^2 + b_n^2)\right\}^{1/2}$.

Finally, one has to bear in mind that spectra resulting from the charge modulation described above do not allow to resolve the question about periodicity of this modulation appearing as CDW and/or EFGW. In general, a combination of CDW and EFGW is a tensorial field in a three dimensional space with six independent components varying across the space – five of them describing EFG and one describing charge (electron) density. Addition of SDW adds another three components to the field as SDW is described locally by the axial vector. Hence, the field having nine components describes variation of the hyperfine Hamiltonian(s) without taking into account possible hyperfine anomaly, the latter being absent for the resonant transition considered here.

The ratio $f/f_0$ of the (average) absorber recoilless fraction $f$ at some temperature to the corresponding recoilless fraction $f_0$ at the reference temperature (here at 4.2 K) is calculated as the ratio of respective products $\Gamma t_A$. It is assumed that recoilless fraction is the same for all resonant atoms. The absorber dimensionless resonant thickness $t_A$ evaluates to $t_A = n_0 \sigma_0 d \, f(\Gamma_0/\Gamma)$ [23]. The symbol $n_0$ stands for the number of resonant nuclei per unit volume within the homogeneous absorber. The symbol $\sigma_0$ denotes resonant cross-section for absorption. The symbol $d$ stands for the absorber thickness along the beam of (collimated) radiation. Note that the product $\sigma_0 \Gamma_0$ does not depend on the total (internal) conversion coefficient, and this product is independent of CDW. A resonant cross-section takes on the form $\sigma_0 = 2\pi \left(\frac{\hbar c}{E_0}\right)^2 \left(\frac{2I_e+1}{2I_g+1}\right)(\Gamma_n/\Gamma_0)$. The symbol $\hbar$ stands for the Planck constant divided by $2\pi$. For a resonant transition from the stable ground nuclear state to the first excited nuclear state (like here) one has $\Gamma_0 = \Gamma_n(1+\alpha_T)$ with the symbol $\alpha_T$ denoting the total internal conversion coefficient and the symbol $0 < \Gamma_n \leq \Gamma_0$ denoting line width for pure single photon radiative transition from the ground to the first excited state. For a transition considered here one has $\alpha_T \approx 9.0$.

In summary, one can state that the parameter $\lambda$ could be measured independently for each spectrum, and it remains pretty constant for a single uninterrupted series of measurements due to the relatively long lifetime of the source. The parameters $\Gamma_S$, $f_S$ and $\Gamma_{exp}$ could be determined from spectrum of the high purity α-Fe foil with the natural isotopic composition for a transition considered here.

## 4. Discussion of results

Figure 1 shows spectra of the parent compound BaFeAs$_2$ at selected temperatures [22] and spectra of the optimally doped superconductor Ba$_{0.6}$K$_{0.4}$Fe$_2$As$_2$ at three selected temperatures. Spin density wave (SDW) order appears below 140 K in the parent compound [22, 24]. Upon potassium doping magnetism gradually disappears with a lowering of the SDW transition temperature [25]. Generally suitable substitution of any element in the '122' parent compounds leads to suppression of SDW and eventual appearance of the superconductivity [26, 27]. The difference in total molar specific heat coefficients $\gamma_s^{tot} - \gamma_p^{tot}$ between superconductor (s) and parent compound (p) versus temperature is also shown together with the electronic specific heat coefficient $\gamma_s^{el}$ of the superconductor versus temperature [18]. Hence, one can conclude that the gap leading to the superconductivity opens at 38 K. On the other hand, the Mössbauer spectra do not show any magnetically split components even at 4.2 K for the superconductor. Nevertheless they are not simple singlets or doublets but exhibit some broad components even at 300 K. Spectra can be fitted as superposition of two doublets, one narrow and the other one very broad. A broad doublet contributes roughly 10 % to the absorption cross-section area. Hence, it is interesting to look at the origin of the broad component. Figure 2 shows spectra at selected temperatures covering transition to the superconducting state and approaching the ground state of the system. The broad feature is still present and one can observe that the spectral shape changes abruptly between 40 K and 38 K (at the superconducting gap opening) and recovers to the previous shape between 28 K and 24 K. This variation affects the broad feature as well. Hence, one has to conclude that the broad feature is not due to a separate phase, and its shape is not governed by the magnetic interactions as it survives till 300 K at least. On the other hand, it is too broad to be accounted for by the variation of the electron density on the resonant nuclei alone. Hence, a distribution of the EFG is essential to explain this phenomenon. Due to the fact, that the broad feature is sensitive to the superconducting transition one has to resort to some kind of EFGW described in the previous section.

Essential parameters derived from data fits to the EFGW model described above are gathered versus temperature $T$ in Figure 3. The average total shift $S$ remains practically unaffected by the transition to the superconducting state indicating that neither average electron density on the iron nuclei nor SOD is sensitive to the transition. On the other hand, the small constant component of the quadrupole splitting $\Delta_0$ shows distinct anomaly below transition. Even larger anomaly is observed for the absorber line width $\Gamma$ in correlation with the anomaly in the dimensionless absorber resonant thickness $t_A$. Somewhat lesser anomaly is seen for the amplitude of EFGW $A$. On the other hand, the parameter $\beta$ responsible for the shape of EFGW shows quite large anomaly. It is interesting to note that all these parameters practically recover to the previous values once the superconducting gap becomes almost fully developed – at about 24 K.

Shapes of EFGW are shown at selected temperatures versus $\mathbf{q} \bullet \mathbf{r}$ in Figure 4. Insets show corresponding distributions $w(\varepsilon - \varepsilon_0) = \rho(\varepsilon - \varepsilon_0)d(\varepsilon - \varepsilon_0) = \rho(\varepsilon - \varepsilon_0)d\varepsilon$ of the parameter $\varepsilon - \varepsilon_0$. The increment $d\varepsilon$ was set as $d\varepsilon = A/63$. Hence, these are weights $w(\varepsilon - \varepsilon_0)$ normalized to unity. The shape of distributions clearly explains why the spectra are seen as composed of the "narrow" and "broad" components. The "sheet" containing large EFG becomes broader and less pronounced in the phase space just below opening of the superconducting gap on the temperature scale. Hence, the spectrum becomes "sharper". A recovery is observed upon further cooling of the sample. The maximum of the quadrupole splitting is reached at the ground state with $\Delta_{\max}$ being about 2.1 mm/s. On the other hand, it amounts to approximately 1.45 mm/s just below transition to the superconducting state. A drop of the maximum quadrupole splitting is about 0.5 mm/s at the gap opening. Such splitting could be observed for highly covalent bonds of iron with the electron(s) located in one of the lobes of the 3d "atomic" state. Such electronic configuration is consistent with the observed total shift of about 0.5 mm/s – at the same temperature. Quite significant EFGW survives till 300 K at least.

Upper part of Figure 5 shows the ratio of the recoilless fractions $f/f_0$ with the symbol $f_0$ denoting recoilless fraction at 4.2 K. There is no sensitivity of the recoilless fraction to the superconducting transition. A small bump just below transition to the superconducting state (smaller than respective error bars) is due to the imperfect approximation of the EFGW shape and approximate treatment of the CDW. One has to note that EFGW and CDW vary significantly just below transition to the superconducting state. The lower part of Figure 5 shows dispersion of CDW $\Delta\rho$ versus temperature. One can see again a distinct anomaly at the transition to the superconducting state. A dispersion of CDW amounts to about 0.5 el./a.u.$^3$ at low temperatures, but out of the anomaly region. It drops to about 0.3 el./a.u.$^3$ within the anomaly region. Hence, a difference is about 0.2 el./a.u.$^3$ on the iron nuclei. It is interesting to note that even at 300 K CDW has quite significant dispersion.

## 5. Conclusions

Optimally doped iron-based superconductor $Ba_{0.6}K_{0.4}Fe_2As_2$ belonging to the '122' family has unusual electronic structure. A modulation of the charge (electron) density develops and it is quite stable versus temperature. A modulation leads to the development of CDW on the iron nuclei, and what is more to the modulation of the EFG on the same nuclei. The latter effect causes much stronger perturbation of the hyperfine interactions on the iron nuclei. It could be accounted for by introducing a new type of the hyperfine interaction modulation called EFGW. The charge modulation is sensitive to the transition between normal and superconducting state. Namely, it is partially suppressed just below opening of the superconducting gap, and it recovers upon fair separation of the bosonic states from the rest of the electronic system. The anomaly in the hyperfine interactions coincides with the peak of the electronic specific heat coefficient $\gamma_s^{el}$. The quadrupole splitting varies among iron nuclei from almost null to about 2.1 mm/s close to the ground state of the system, while the electron density on the iron nuclei varies by about 0.5 el./a.u.$^3$ at low temperatures. Hence, one can conclude that covalent bonds between iron and arsenic play important role in this otherwise metallic system. On the other hand, a distribution of the "covalent" electrons is strongly perturbed by the itinerant electrons forming Cooper pairs. Formation of CDW (nematic order) has been discussed in the iron-based superconductors, but in a very vague fashion and without realization that it affects electronic states with the non-zero angular momentum leading to EFGW [28]. Dynamic properties of the iron nuclei (recoilless fraction and SOD) seem unaffected by a transition to the superconducting state.

**Acknowledgments**

Professor Józef Spałek (Institute of Physics, Jagiellonian University and Faculty of Physics and Applied Computer Science, AGH University of Science and Technology, Kraków, Poland) and Dr. John W. Loram (Cavendish Laboratory, University of Cambridge, U. K.) are warmly thanked for helpful comments.

This work was supported by the National Science Center of Poland, Grant DEC-2011/03/B/ST3/00446. J. K. acknowledges support by the SNSF (Project No. 140760) and FP7 Super-Iron Project.

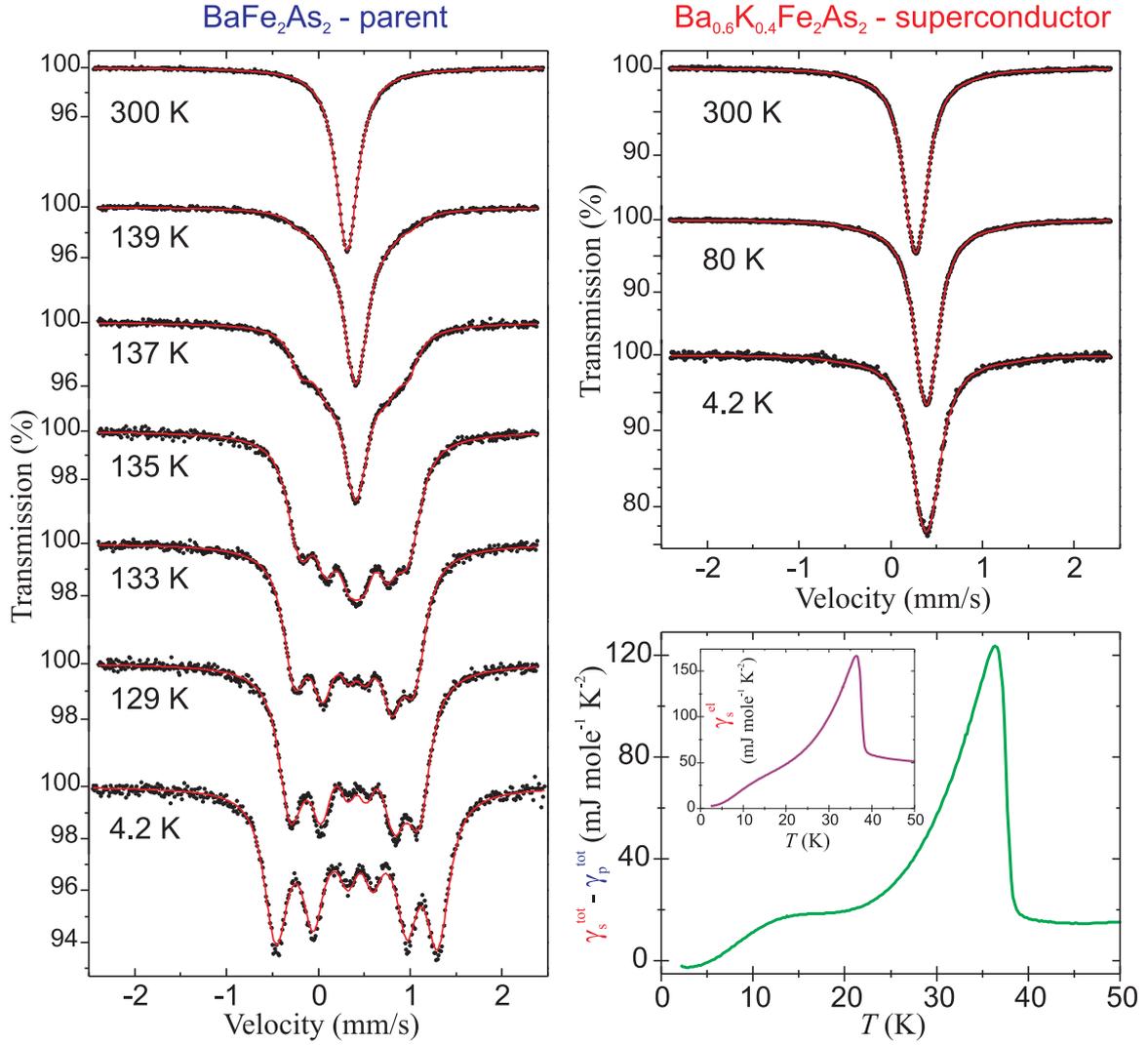

**Figure 1** $^{57}$Fe Mössbauer spectra versus temperature for the parent compound BaFe$_2$As$_2$ [22] and the Ba$_{0.6}$K$_{0.4}$Fe$_2$As$_2$ superconductor. Solid lines are results of the fit to data. The difference in total molar specific heat coefficients $\gamma_s^{tot} - \gamma_p^{tot}$ between superconductor (s) and parent compound (p) versus temperature is also shown with $\gamma^{tot} = C^{tot}/T$. The symbol $C^{tot}$ stands for the total molar heat capacity and $T$ is the temperature. The inset shows the electronic specific heat coefficient $\gamma_s^{el} = C^{el}/T$ of the superconductor versus temperature [18]. The symbol $C^{el}$ stands for the electronic molar heat capacity. All heat capacities were measured in the zero applied magnetic fields and under ambient pressure.

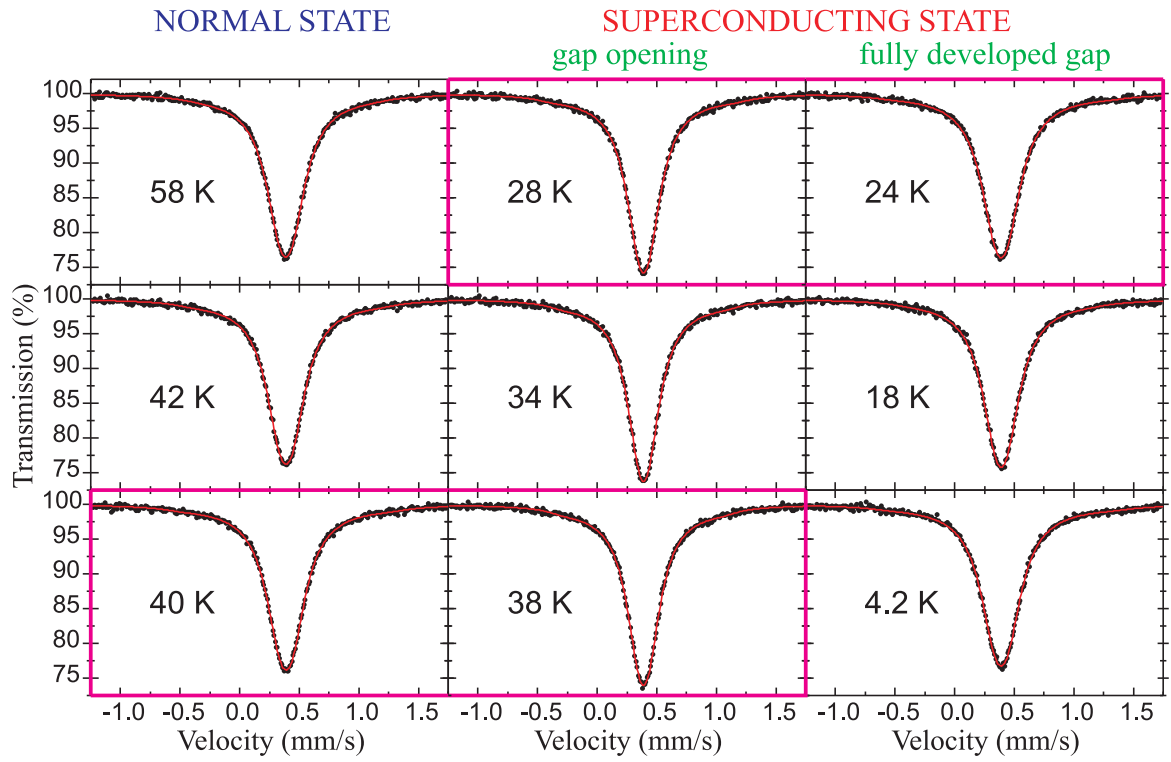

**Figure 2** Selected Mössbauer spectra of the $Ba_{0.6}K_{0.4}Fe_2As_2$ ( $T_{sc} = 38\,K$ ) across the transition to the superconducting state. Solid lines are results of the fit to data. Note the abrupt changes in the regions 40 K - 38 K and 28 K - 24 K.

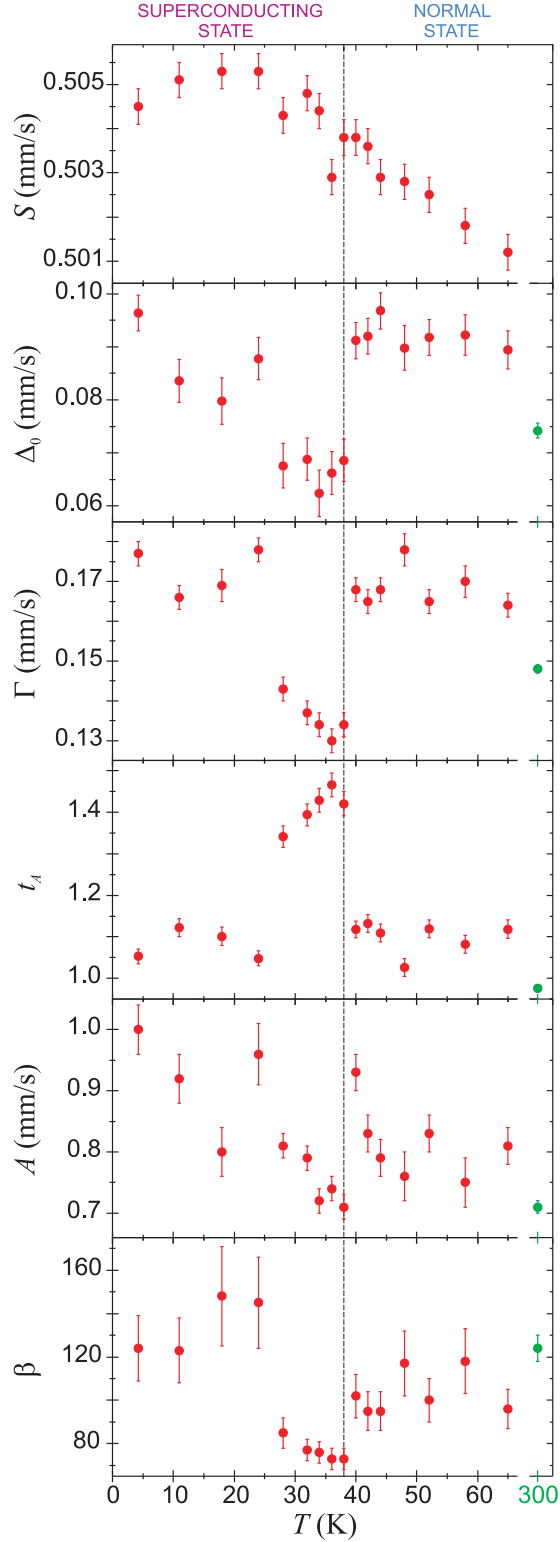

**Figure 3** Essential parameters derived from the Mössbauer spectra of the $Ba_{0.6}K_{0.4}Fe_2As_2$ are plotted versus temperature $T$. Symbol $S$ stands for the total spectrum shift versus room temperature α-Fe and symbol $\Delta_0$ denotes constant component of the quadrupole splitting. Symbol $\Gamma$ stands for the absorber line width, and $t_A$ denotes dimensionless absorber resonant thickness. Symbol $A$ stands for the amplitude of EFGW, while $\beta$ denotes the shape parameter of EFGW. Dashed vertical line marks transition between normal and superconducting states. The total shift $S$ amounts to 0.3883(2) mm/s at 300 K.

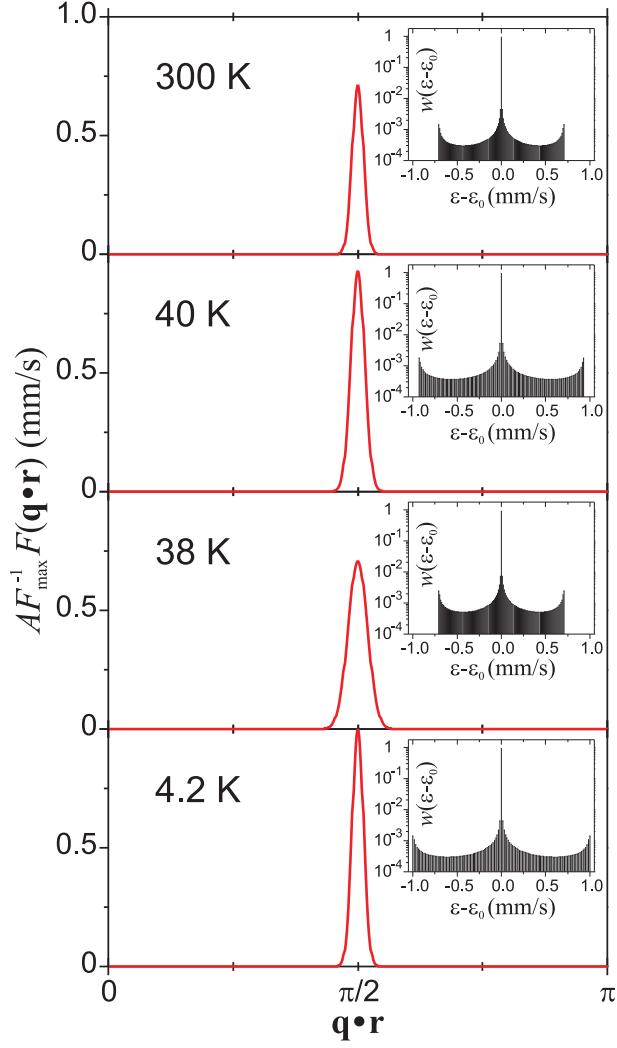

**Figure 4** Shape of the EFGW is shown for selected temperatures versus phase $\mathbf{q}\cdot\mathbf{r}$. Insets show corresponding normalized weights $w(\varepsilon-\varepsilon_0)$ of the quadrupole coupling constant $\varepsilon-\varepsilon_0$.

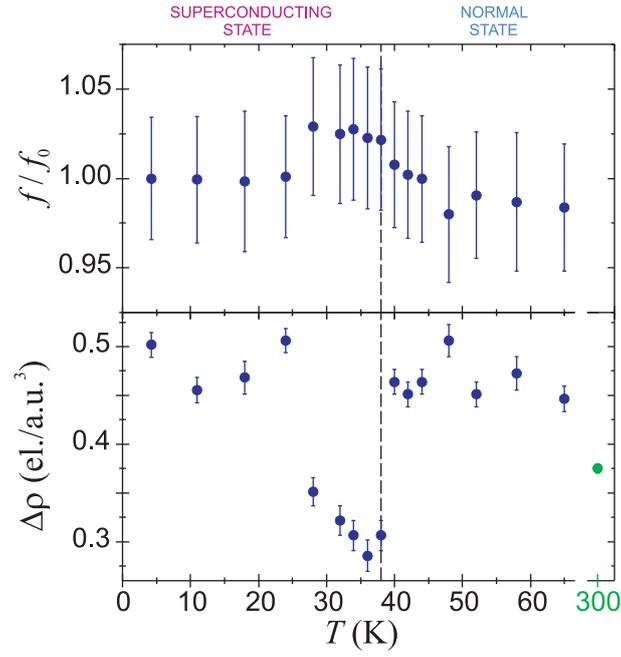

**Figure 5** Ratio of the recoilless fractions $f/f_0$ is plotted versus temperature in the upper part with the symbol $f_0$ denoting recoilless fraction at 4.2 K. The ratio $f/f_0$ amounts to 0.78(2) at 300 K. Lower part shows dispersion of CDW $\Delta\rho$ versus temperature. Dashed vertical line marks transition between normal and superconducting state.